\documentstyle[twocolumn,aps]{revtex}
\input{epsf}

\begin{document}

%set sloppy attitude to line breaks
%\sloppy
%\setcounter{page}{0}
%\thispagestyle{empty}

%\begin{flushright}
%nucl-th/9802034
%\end{flushright}
%\vspace{3.5cm}

\title{FLASH OF PHOTONS FROM THE EARLY STAGE OF HEAVY-ION COLLISIONS}

\author{Dinesh K. Srivastava$^1$ and Klaus Geiger$^2$}

\address{$^1$ Variable Energy Cyclotron Centre, 1/AF Bidhan Nagar, Calcutta 700 064, India}
\address{$^2$ Physics Department, Brookhaven National Laboratory, Upton, N. Y. 11973, U. S. A.}

\date{\today}

\maketitle

\begin{abstract}
The dynamics of partonic cascades may be an important aspect
for particle production in
relativistic collisions of nuclei at CERN SPS  and BNL RHIC energies.
Within the {\it Parton-Cascade Model}, we estimate the production of single 
photons from such cascades due to  scattering of quarks and gluons
$q~g~\rightarrow~ q~\gamma$, quark-antiquark annihilation 
$q~\bar{q}~\rightarrow~g~\gamma$ or $\gamma\gamma$, 
and  from electromagnetic
brems-strahlung of quarks $q~\rightarrow~ q~\gamma$.
We find that the latter QED branching process plays the dominant role
for photon production, similarly as
the QCD branchings $q\rightarrow qg$ and  $g\rightarrow gg$ 
play a crucial role for parton multiplication.
We conclude therefore that photons accompanying the
parton cascade evolution during the early stage of heavy-ion collisions
shed  light on the formation of a partonic plasma.
\end{abstract}


%
%\newpage

Photons have remained one of the most
effective probes of every kind of terrestrial or celestial matter over the
ages. Thus, it is only befitting that the speculation of the
formation of  deconfined strongly interacting matter - some form of
the notorious quark-gluon plasma - in relativistic heavy-ion collisions,
was soon followed by a suggestion \cite{shu78} that it should
be accompanied by characteristic production of photons. 
The effectiveness of photons in probing the history of such 
a hot and dense matter stems from the fact that, after production, 
they leave the system without any further interaction and thus carry 
unscathed information about the circumstances of their birth.
This is a very important consideration indeed, as the formation of 
a partonic plasma is likely to proceed from a hard-scattering of initial
partons, through a pre-equilibrium stage, to perhaps a thermally and 
chemically equilibrated state of hot and dense partonic matter.
In this letter we concentrate on photons coming from the
early partonic stage in such collisions.
We will not consider other sources of photons, e.g., those
that accompany the late hadronic stage, after the partons have hadronized.

During the  partonic stage, photons emerge from two different mechanisms:
firstly, from collisions between  partons, 
i.e., Compton scattering  $qg \rightarrow q\gamma$ of quarks and gluons
and annihilation $q\bar q\rightarrow g\gamma$, $q\bar q\rightarrow \gamma\gamma$
of quarks and antiquarks;
secondly, from radiation of excited partons, i.e.
electromagnetic brems-strahlung $q\rightarrow q \gamma$ of time-like 
cascades initiated by quarks.
Whereas the former mechanism has been studied in various contexts \cite{scat},
the latter source of photons is less explored \cite{sjoestrand}, 
although, as we shall show, it is potentially much richer both in magnitude and complexity.

A dynamical description of  relativistic heavy-ion collisions
has been developed in the {\it Parton Cascade Model} (PCM) \cite{pcm}.
The PCM is based on the parton picture of hadronic
interactions and describes the nuclear dynamics in terms of
the interaction of quarks and gluons within perturbative quantum
chromodynamics, embedded in the framework of relativistic
quantum kinetics \cite{ms3942}. The time evolution of the system is followed by
Monte Carlo simulation of a set of equations describing space-time development
and renormalization-group evolution of the particles.
The procedure, implemented in a computer code VNI~\cite{vni},
traces the dynamic evolution of
partons scattering, radiating, fusing, and eventually clustering to
pre-hadronic states that then produce the final-state hadron yield.
VNI, the Monte Carlo implementation of the PCM, has been adjusted on the
basis of experimental data from $e^+e^-$ annihilation and $pp$ ($p\bar{p}$)
collisions.
In the past, the PCM has been  used to provide insight \cite{pcm} into 
conditions likely to be achieved at RHIC and LHC energies $\sqrt{s} = 200$ A GeV, 
respectively $\sqrt{s} = 5$ A TeV.
Only very recently, it has been found~\cite{gs} to provide 
as well a reasonable description to a large body of particle spectra 
from $Pb+Pb$ and $S+S$ collisions at CERN-SPS energy $\sqrt{s} \simeq 18$ A GeV, 
in contrast to the belief that the PCM could not be applied at energies
$\sqrt{s}  \, \lower3pt\hbox{$\buildrel <\over\sim$}\,100$ A GeV.
A valuable advantage of the PCM is that it is free 
of assumptions about thermal or chemical equilibrium conditions, since
the space-time evolution of the matter is traced causally from the moment
of collision on, so that at any point the state of the matter is 
unambiguously determined by the preceding space-time history.

Prompt photons 
from both parton collisions and brems-strahlung are ideally suited to test 
the evolution of the partonic matter as described by the PCM. 
They would accompany the early hard scatterings,
the approach to thermalization and chemical equilibration if these are 
achieved at all.

We first discuss
the aspect of {\it collisional photon production} from parton
collisions. This is straightforward, and is included in the PCM 
in terms of the elementary $2\rightarrow 2$ processes which yield photons,
that is,
the annihilation and Compton processes,
$
q\bar{q}   \rightarrow  g \gamma 
,\;
q\bar{q}   \rightarrow  \gamma \gamma 
,\;
qg \rightarrow  q \gamma~
$,
the Born cross-sections
of which are well-known \cite{xsections}. In the PCM approach we treat these
processes within perturbative QCD if the momentum 
$q_\perp$ transferred in the parton collision is larger than some 
cut-off $q_\perp^0$ \cite{pcm}. 
(This cut-off is introduced 
in the c.m. frame of the colliding partons, and thus we may
have contributions even for smaller transverse momentum in the nucleus-nucleus 
c.m. frame.)

Next we turn to
the aspect of {\it radiative photon emission} from parton
showers initiated by collisions, which is a more complex issue.
There are some important and interesting differences
between a branching leading to production
of photons as compared to gluons, 
which have been pointed out in detail by Sj\"ostrand \cite{sjoestrand}. 
%\begin{description}
%\item[(i)]

\noindent {\bf (i)}
Consider an energetic quark produced by a hard scattering,
which can radiate gluons and photons competitively.
The branchings $q\rightarrow q g$
and $q\rightarrow q \gamma$ appear in the PCM on an equal footing and as
competing processes with similar structures. The probability, for a quark
to branch at some given virtuality scale $Q^2$, with the daughter quark
retaining a fraction $z$ of the energy of the mother quark, is given by: 
\begin{equation}
d{\cal{P}}=\left( \frac{\alpha_s}{2\pi}C_F +\frac{\alpha_{\rm {em}}}{2\pi}e_q^2
         \right )  \frac{dQ^2}{Q^2} \frac{1+z^2}{1-z} dz
\end{equation}
where the first term corresponds to gluon emission and the second to photon
emission. Thus, the relative probability for the two processes is,
\begin{equation}
\frac{ {\cal {P}}_{q\rightarrow q\gamma} }
     { {\cal {P}}_{q\rightarrow q g} }
\;\propto\;\frac{\alpha_{\rm{em}} \langle e_q^2 \rangle }{\alpha_s C_F}
\;\simeq \;\frac{1}{200}
\;,
\end{equation}
using
$\alpha_{\rm{em}}=1/137$, $\alpha_s = 0.25$, $ \langle e_q^2 \rangle = 0.22$
and $C_F = 4/3$.
Thus,  we notice that the emission of photons is strongly affected by the
perturbative QCD effects. This does not mean, though, that
we can simulate emission of photons in a QCD shower by simply
replacing the strong coupling constant $\alpha_s$ with the
electromagnetic $\alpha_{\rm{em}}$ and the QCD color Casimir factor $C_F=4/3$
by $e_q^2$.  One has to  keep in mind that
the gluon,  thus emitted, may branch further either as,
$g\rightarrow gg$ or as $g \rightarrow q\bar{q}$,
implying that the emitted gluon has an effective non-zero mass.
As the corresponding probability for the photon to branch into a quark
or a lepton pair is very small, this process is neglected and the
photon is taken to have a zero mass. 
 (However, if we wish to study the dilepton
production from the collision, this may become an important
contribution~\cite{GK}).

%\item[(ii)]

\noindent {\bf (ii)}
The radiation of gluons
from the quarks is subject to soft-gluon interference which is accounted
for by imposing an angular ordering of the emitted gluons. This is
not so for the emitted photons. To recognize this aspect,
consider a quark which has already radiated a number of hard gluons.
The probability to radiate and additional softer gluon will get contributions
from each of the existing partons which may further branch  as 
$q\rightarrow qg$ or $g \rightarrow gg$. It is well-known \cite{bassetto}
that if such a soft gluon is radiated at a large angle with respect to all
the other partons and one would add the individual contributions incoherently,
then the emission rate would be overestimated, as the interference is
destructive. This happens because a  soft gluon of a long wavelength 
is not able to
resolve the individual color charges and observes only the net charge. 
The probabilistic picture of PCM is then recovered by demanding that
emissions are ordered in terms of decreasing opening angle 
between the two daughter partons at each branching, i.e., restricting the
phase-space allowed for the successive branchings. The photons, on the
other hand, do not carry any charge and only the quarks radiate.
Thus photons are not subject to angular ordering. 
Pictorially,
the branching structure in QCD is `pine-tree' like, whilst in QED
it is `oak-tree' like.

%\item[(iii)]

\noindent {\bf (iii)}
The parton emission probabilities in the QCD showers contain
soft and collinear singularities,  which are regulated by introducing
a cut-off scale $\mu_0$. This regularization procedure
implies  effective masses for quarks and gluons,
\begin{equation}
m_{\rm{eff}}^{(q)}=\sqrt{\frac{\mu_0^2}{4} +m_q^2}
\;,\;\;\;\;\;\;\;
m_{\rm{eff}}^{(g)}=\frac{\mu_0}{2}
\;,
\label{mu0}
\end{equation}
where $m_q$ is the current quark mass. Thus the gluons cannot branch
unless their mass is more than $2 m_{\rm{eff}}^{(g)} = \mu_0$,
while  quarks cannot  branch unless their mass is more than
$m_{\rm{eff}}^{(q)}+m_{\rm{eff}}^{(g)}$. 
An appropriate value for $\mu_0$ is about 1 GeV \cite{pcm}; a larger value is
not favored by the data, and a smaller value will cause the perturbative
expression to blow up.  These arguments, however,  do not apply for photon
emission, since QED perturbation theory does not break-down and photons
are not affected by confinement forces. Thus, in principle quarks
can go on emitting photons till their masses reduce to current quark masses
(or, in a dense matter environment, to the corresponding `in-medium'
Debye mass).
It has also been suggested that if the confinement
forces screen the bare quarks,  the effective cut-off can be
of the order of a GeV. These arguments suggest that
we can choose the cut-off scale $\mu_0$ in (\ref{mu0}) separately for
the emission of photons and that the study of photon emission can provide
valuable insight about confinement at work by comparing the
characteristics of gluon versus photon radiation from quarks.
%\end{description}
\smallskip

We now turn to our results which we present
in four examples: $S+S$ ($\sqrt{s} = 20$ A GeV) and $Pb+Pb$ collisions
($\sqrt{s} \simeq 18$ A GeV) at CERN-SPS, as well as  
$S+S$ and $Au+Au$ collisions at RHIC energy $\sqrt{s} = 200$ A GeV.
For the following, it is useful to keep in mind, that the initial partonic composition
of the colliding nuclei is very different for the two energies:
at SPS, by far the main component are the quarks,
with relative proportions $N_q : N_g \approx  1 : 0.4$,
whereas at RHIC, the gluons play the
dominant role, with $N_q : N_g\approx 1 : 1.9$.
\smallskip

In Fig.~1 we have plotted the production of single photons from such a
partonic matter in the central rapidity region for $Pb+Pb$ system at SPS
energy.  The dot-dashed histogram shows the contribution of Compton and
annihilation processes mentioned above. The dashed and the solid
histograms show the total contributions (i.e., including the branchings
$q\rightarrow q\gamma$) when the mass scale $\mu_0$ in (\ref{mu0})
for the photon production is taken respectively as 0.01 and 1 GeV. 
We see that prompt photons from the quark branching completely 
dominate the yield for
$p_\perp \leq 3$ GeV, whereas at larger transverse momenta the
photons coming from the collision processes dominate.  The reduction
of $\mu_0$ for the $q \rightarrow q\gamma$ branching
is seen to enhance the production of photons have lower transverse
momenta as one  expects.
We have also shown the production of single photons from $pp$ 
collisions for $\sqrt{s}\approx$ 24  GeV,
obtained by WA70~\cite{wa70}, NA24~\cite{na24}, and UA6~\cite{ua6}
collaborations scaled by the nuclear overlap for zero impact parameter for
the collision of lead nuclei. The solid curve gives the
perturbative QCD results~\cite{jean} for the $pp$ collisions
scaled similarly. The dashed curve is a direct extrapolation
of these results to lower $p_\perp$.
\medskip

%%%%%%%%%%%%%%%%%%%%%%%%%%%%%%%%  Fig. 1   %%%%%%%%%%%%%%%%%%%%%%%%%%%%
\begin{figure}[htb]
\epsfxsize=200pt
\rightline{ \epsfbox{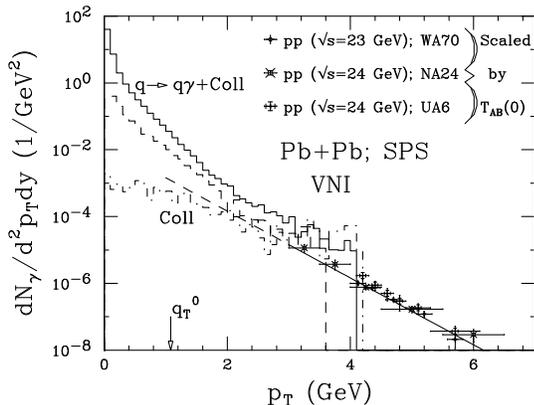} $\;\;\;\;\;\;\;\;\;$  }
%\vspace{-0.8cm}
\caption{ 
 The radiation of prompt photons from                              
 partonic matter in central collision of Pb (156  GeV/nucleon)+Pb
  nuclei at CERN SPS. The dot-dashed histogram gives the contribution
  of only the collision processes. The dashed and the solid 
  histograms give the contribution of the collision plus branchings
  when the $\mu_0$ for the $q~\rightarrow~q\gamma$ branching is taken
  as 1 and 0.01 GeV respectively. The $pp$ data at $\sqrt{s}\approx
  24$ GeV scaled by the nuclear overlap function for the central
  collision of lead nuclei and the corresponding QCD predictions
  (solid curve, arbitrarily extended to lower $p_T$) also shown
  for a comparison. $q_T^0$ denotes the  transverse
  momentum cut-off above which the hard scatterings are
  included in the PCM.
  }
\end{figure}
%%%%%%%%%%%%%%%%%%%%%%%%%%%%%%%%%%%%%%%%%%%%%%%%%%%%%%%%%%%%%%%%%%%%%%%

In Fig.~2 we have plotted the transverse momenta of the single photons
in several rapidity bins for $Pb+Pb$ and $S+S$ systems at SPS energy.
We see that the transverse spectra scale reasonably well with the
ratio of the nuclear overlap for central collisions for the
two systems $T_{\rm{PbPb}}/T_{\rm{SS}}\approx$ 15.4, which is indicative of
the origin of these photons basically from a collision mechanism. 
The slight deviation from this scaling seen at lower
$p_\perp$ results in a $\approx$ 20\% increase in the integrated yield
at central
rapidities. This is a good measure of the multiple
scatterings in the PCM.  In fact we have found that the number of hard
scatterings in the $Pb+Pb$ system is $\approx$ 17 times more
than that for the $S+S$ system, which also essentially determines the
ratio of the number of the photons produced in the two cases.
We also note that the inverse slope of the $p_\perp$ distribution
decreases at larger rapidities, which is suggestive of the fact that
the densest partonic system is formed at central rapidities.
%\medskip

Finally in Fig.~3 we have plotted our results for $S+S$ and $Au+Au$ systems
at RHIC energy in the same fashion as Fig.~2. We see that the
inverse slope of the $p_\perp$ distribution is now larger and 
drops only marginally at larger  rapidities,
indicating that the  partonic system is now more dense and
spread over a larger range of rapidity.
Even though the $p_\perp$ distribution of the photons is seen to
roughly scale with the ratio of the nuclear overlap functions
for central collisions $T_{\rm{AuAu}}/T_{\rm{SS}}\approx$ 14.2,
the integrated yield of photons for the $Au+Au$ is seen to be only about 
12 times  that for the $S+S$ system at the RHIC energy. We have
again checked that the number of hard scatterings for the
$Au+Au$ system is also only about 12 times that for the $S+S$
system. 

%%%%%%%%%%%%%%%%%%%%%%%%%%%%%%%%  Fig. 2   %%%%%%%%%%%%%%%%%%%%%%%%%%%%
\begin{figure}[htb]
\epsfxsize=200pt
\rightline{ \epsfbox{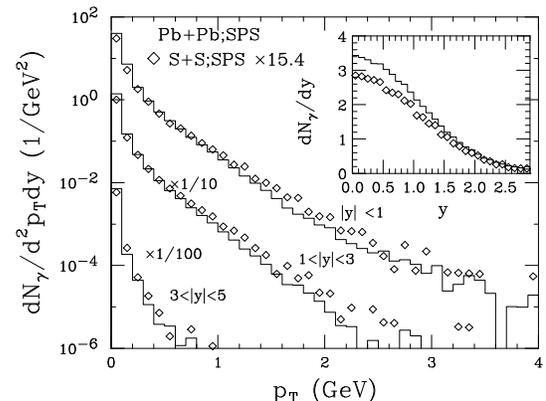} $\;\;\;\;\;\;\;\;\;$  }
%\vspace{-0.8cm}
\caption{ 
The radiation of single photons from $S+S$(200~GeV/A) and
$Pb+Pb$(158 GeV/A) collisions in different rapidity bins.
The inset shows the rapidity distribution of the radiated photons.
The results for the $S+S$ collisions have been scaled by the
ratio $T_{\rm{PbPb}}/T_{\rm{SS}}\approx~15.4$ for central
collisions.
$\mu_0$ for the quark branching $q~\rightarrow~q\gamma$
is taken as 0.01 GeV.
}
\end{figure}
%%%%%%%%%%%%%%%%%%%%%%%%%%%%%%%%%%%%%%%%%%%%%%%%%%%%%%%%%%%%%%%%%%%%%%%

%%%%%%%%%%%%%%%%%%%%%%%%%%%%%%%%  Fig. 3   %%%%%%%%%%%%%%%%%%%%%%%%%%%%
\begin{figure}[htb]
\epsfxsize=200pt
\rightline{ \epsfbox{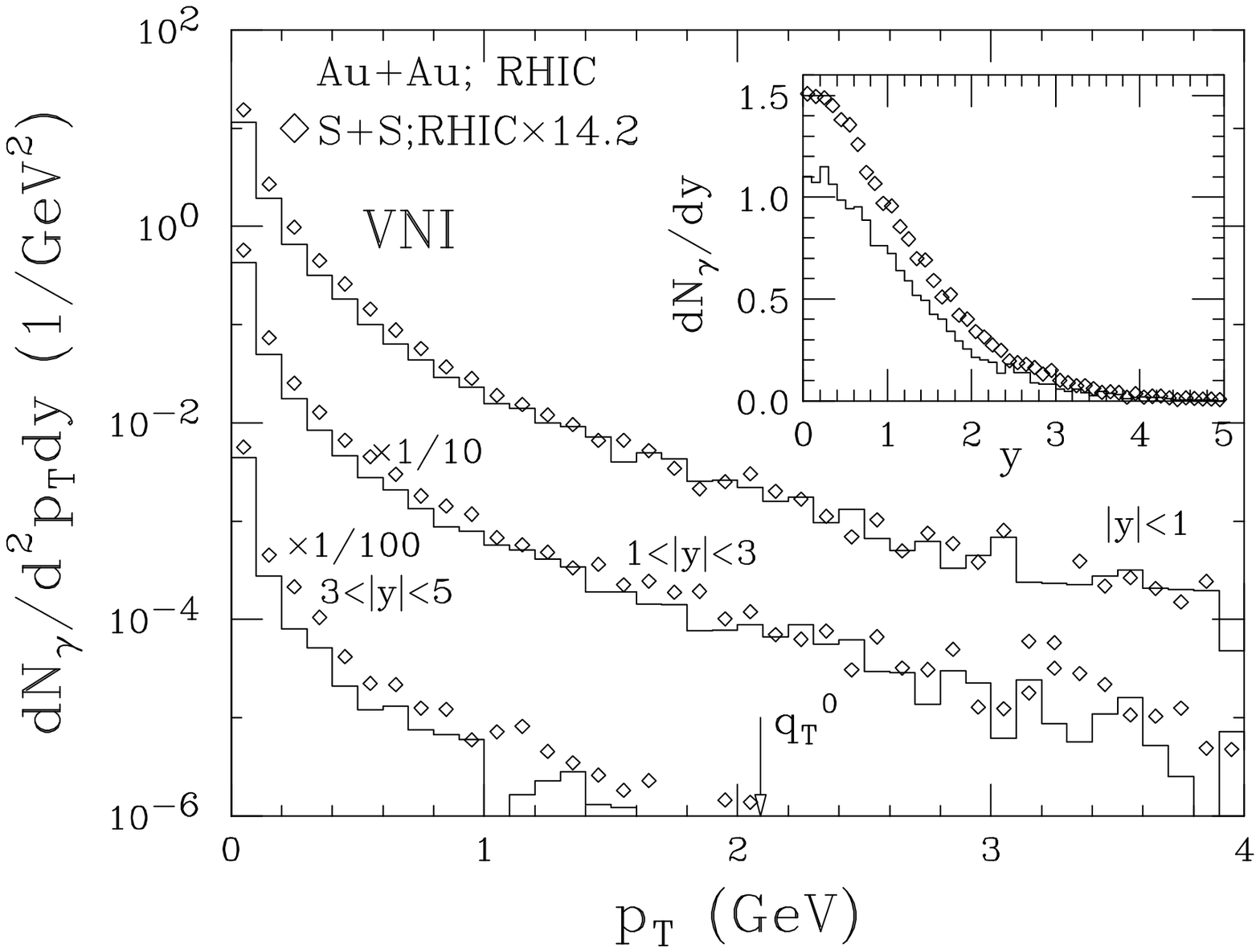} $\;\;\;\;\;\;\;\;\;$  }
%\vspace{-0.8cm}
\caption{ 
The radiation of single photons from $S+S$ and
$Au+Au$ collisions in different rapidity bins at RHIC energy 
$\sqrt{s}= 200$ A GeV.
The inset shows the rapidity distribution of the radiated photons.
The results for the $S+S$ collisions have been scaled by the
ratio $T_{\rm{AuAu}}/T_{\rm{SS}}\approx~14.2$ for central collisions.
$\mu_0$ for the quark branching $q~\rightarrow~q\gamma$
is taken as 0.01 GeV.
               }
\end{figure}
%%%%%%%%%%%%%%%%%%%%%%%%%%%%%%%%%%%%%%%%%%%%%%%%%%%%%%%%%%%%%%%%%%%%%%%

This contrasting behavior at SPS and RHIC energies exhibited in the figures
can be understood as follows. At the SPS,
the partonic system can reach energy densities up to 5 GeV/fm$^3$
\cite{gs}, and multiple scatterings become important, especially for heavier
colliding nuclei. At RHIC, the energy density can easily become 
twice as large, and the Landau-Pomeranchuk suppression may play a counteracting role.
In the PCM this effect is mimicked, rather crudely,
by inhibiting a new scattering
of partons till the passage of their
formation time after a given scattering.
In a future publication~\cite{multscat} we shall report
on a possibility of seeing these competitive mechanisms
at work by comparing results at different impact parameters
for the same colliding nuclei, or for zero impact paramenter for 
different colliding nuclei. 
\medskip

Some other observations are worthwhile commenting on:
Firstly, recall such branchings of the partons produced in hard collisions
correspond to a next-to-leading-order correction in $\alpha_s$.
These are known to be  considerably enhanced for collinear
emissions. The parton shower mechanism incorporated in the PCM
amounts to including these enhanced contributions to all orders,
instead of including all the terms for a given order~\cite{rkellis}.
It may also be added that the first-order corrections to the
Compton and annihilation processes in the plasma have
been studied by a number of authors~\cite{pradip}; however
in the plasma,  $\langle q_\perp^2\rangle  \approx 4T^2$, 
thus their contribution is limited to very low transverse momentum. 
Clearly, $\langle q_\perp^2 \rangle$ is much larger in the early
hard scatterings,  and thus the radiations from the emerging
partons are much more intense and also populate higher transverse momenta,
as seen in the present work.
The large yield of photons from the branching of energetic quarks
preceding the formation of dense partonic matter opens an interesting
possibility to look for a similar contribution to dilepton (virtual
photon) production in such collisions.
\bigskip

We conclude that the formation of a hot and dense partonic
system in relativistic heavy-ion collisions may be preceded
by a strong flash of photons following the early hard scatterings.
Their yield will, among several other interesting aspects, also
shed light on the extent of multiple scattering encountered
in these collisions \cite{multscat}.
However, we stress that  we have only included the contribution of
photons from the partonic interactions in this work. It is
quite likely that the hadrons produced at the end will also
interact and produce photons, as has been 
studied  recently \cite{crs}. A comparison of
those results with the present work
shows that at the SPS energy the emission from the early
hard partonic scatterings is of the same order as 
the photon production  from later hadronic reactions, for 
$p_\perp \leq$ 2--3 GeV,
and dominates considerably over the same at higher transverse momenta. 
\bigskip

%\section*{ACKNOWLEDGEMENTS}
%{\bf ACKNOWLEDGEMENTS:}
One of us (DKS) would like to acknowledge the hospitality of
Brookhaven National Laboratory, where most of this work was done.
We also thank Dr Bikash Sinha for useful comments.
This work was supported in part by the D.O.E. under contract no.
DE-AC02-76H00016.

%\newpage

\end{document}